\documentstyle[12pt]{article}
\input epsf.sty

\newtheorem{Theorem}{Theorem}
\newtheorem{Lemma}{Lemma}
\newtheorem{Claim}{Claim}
\newtheorem{Corollary}{Corollary}

\def\bbbn{{\rm I\!N}}

\def\bbbc{{\mathchoice {\setbox0=\hbox{$\displaystyle\rm C$}\hbox{\hbox
to0pt{\kern0.4\wd0\vrule height0.9\ht0\hss}\box0}}
{\setbox0=\hbox{$\textstyle\rm C$}\hbox{\hbox
to0pt{\kern0.4\wd0\vrule height0.9\ht0\hss}\box0}}
{\setbox0=\hbox{$\scriptstyle\rm C$}\hbox{\hbox
to0pt{\kern0.4\wd0\vrule height0.9\ht0\hss}\box0}}
{\setbox0=\hbox{$\scriptscriptstyle\rm C$}\hbox{\hbox
to0pt{\kern0.4\wd0\vrule height0.9\ht0\hss}\box0}}}}
\def\cents{\hbox{\rm\rlap/c}}

\begin{document}

\thispagestyle{empty}
\bibliographystyle{latex8}

\date{}

\title{\bf 1-way quantum finite automata:
strengths, weaknesses and generalizations}

\author{Andris Ambainis\thanks{Address: Computer Science Division, 
       University of California, Berkeley, CA 94720,
       e-mail: {\tt ambainis@cs.berkeley.edu}.
       Supported by Berkeley Fellowship for Graduate Studies.
       Part of this work done during 1998 I.S.I.-Elsag Bailey
       research meeting on quantum computing.}\\
       UC Berkeley
       \and
       R\=usi\c n\v s Freivalds\thanks{Address: 
       Institute of Mathematics and Computer
       Science, University of Latvia, Raina bulv. 29, Riga, LV-1459, Latvia,
       e-mail: {\tt rusins@cclu.lv}.
       Supported by Latvia Science Council Grant 96.0282}\\
       University of Latvia}

\maketitle

\begin{abstract}
We study 1-way quantum finite automata (QFAs).
First, we compare them with their classical counterparts.
We show that, if an automaton is required to give the correct
answer with a large probability (greater than 7/9), then 
any 1-way QFA can be simulated by a 1-way reversible automaton.
However, quantum automata giving the correct answer with 
smaller probabilities are more powerful than reversible automata.

Second, we show that 1-way QFAs can be very space-efficient.
%Namely, 
We construct a 1-way QFA that is exponentially smaller
than any equivalent classical (even randomized) finite automaton.
We think that this construction may be useful for design of 
other space-efficient quantum algorithms.

Third, we consider several generalizations of 1-way QFAs.
Here, our goal is to find a model which is more powerful
than 1-way QFAs keeping the quantum part as simple as possible. 
\end{abstract}

\maketitle

\section{Introduction}

It is quite possible that the first implementations
of quantum computers will not be fully quantum mechanical.
Instead, they may have two parts: a quantum part and 
a classical part with a communication between two parts. 
In this case, the quantum part will be
considerably more expensive than the classical part.
Therefore, it will be useful to make the quantum part
as small as possible even if it leads to some (reasonable)
increases in the size of the classical part.
This motivates the study of systems with a small quantum 
mechanical part. 

Quantum finite automata (QFA) is a
theoretical model for such systems.
\cite{KW} introduced both 1-way and 2-way QFAs,
with emphasis on 2-way automata because they are more
powerful. However, the model of 2-way QFAs is not quite consistent
with the idea of a system with a small quantum mechanical part.
\cite{KW} allows superpositions where different parts of superposition
have the head of QFA at different locations.
(Even more, using such superpositions was the main
idea in the proof that 2-way QFAs are more powerful than
classical finite automata.)
This means that the position of the head must 
be encoded into quantum state.
Hence, the number of quantum states necessary to
implement a 2-way QFA is not a constant but grows when
the size of the input increases.
%If we try to implement a 2-way QFA, we will have to 
%represent the whole input in a quantum form.
%(If we store it in classical form, then a QFA trying to read
%a symbol from the input will show to the environment which
%symbol is being read and the whole superposition will decoher.)
%To do it, we will need the amount of quantum bits
%linear in the size of input. This may be very expensive.
This also makes state transformations more complicated
(and more difficult to implement).

Hence, we think that more attention should be given
to the study of simpler models like 1-way QFAs. 
A 1-way quantum automaton is a very reasonable 
model of computation and it is easy to see how it can
be implemented. The finite dimensional state-space 
of a QFA corresponds to a system with finitely many particles. 
Each letter has a corresponding unitary transformation 
on the state-space. A classical device can read 
symbols from the input and apply the corresponding
transformations to the quantum mechanical part.
%In fact, several practical experiments in quantum 
%computing can be viewed as building such systems.

Results about 1-way QFAs in \cite{KW} were quite 
pessimistic. It was shown that the class of languages
recognized by 1-way QFAs is a proper subset of
regular languages. We continue the investigation
of 1-way QFAs and show that, despite being limited
in some situations, they perform well in 
other situations.

Our first results consider relations between 1-way QFAs
and 1-way reversible automata. 
Clearly, a 1-way reversible automaton is a special
case of a QFA and, therefore, cannot recognize all regular
languages. It is a natural question whether 1-way QFAs 
are more powerful than 1-way reversible automata.
Interestingly, the answer depends on the accepting probability
of a QFA. If a QFA gives a correct answer with a large probability
(greater than 7/9), it can be replaced by a 1-way reversible 
automaton. However, this is not true for 0.68... and smaller
probabilities. %We also introduce a model of reversible automata
%with probabilistic choices which is intermediate
%between reversible automata and QFAs and investigate it.

Then, we show that QFAs can be much more space-efficient
than deterministic and even probabilistic finite automata.
Namely, there is a 1-way QFA that can check whether the number
of letters received from the input is divisible by a prime $p$ 
with only $O(\log p)$ states (this is equivalent to $\log \log p$ bits
of memory). Any deterministic or probabilistic finite automaton
needs $p$ states ($\log p$ bits of memory). We think that this
space-efficient quantum algorithm may be interesting for
design of other quantum algoritms as well.

Finally, we consider modifications of 2-way quantum automata
where the head is always at the same position for all parts
of superposition. Modified 2-way QFAs can be implemented with a 
quantum system of constant size. Several modifications 
are proposed. %All regular languages can be recognized by
%modified 2-way QFAs. 
In one of our models (1-way QFAs with
a probabilistic preprocessing), some non-regular languages
can be recognized. %For other models, a similar question remains 
%open.

\section{Definitions}
\label{S2}

\subsection{Quantum finite automata}

We consider 1-way quantum finite automata (QFA) 
as defined in \cite{KW}. Namely, a 1-way QFA is a tuple 
$M=(Q, \Sigma, \delta, q_0, Q_{acc}, Q_{rej})$
where $Q$ is a finite set of states, $\Sigma$ is
an input alphabet, $\delta$ is a transition function, 
$q_0\in Q$ is a starting state and
$Q_{acc}\subset Q$ and $Q_{rej}\subset Q$ are sets of accepting and
rejecting states. The states in $Q_{acc}$ and $Q_{rej}$ are 
called {\em halting states} and the states
in $Q_{non}=Q-(Q_{acc}\cup Q_{rej})$ are called 
{\em non-halting states}.
$\cents$ and $\$$ are symbols that do not belong to $\Sigma$.
We use $\cents$ and $\$$ as the left and the right endmarker, respectively.
The {\em working alphabet} of $M$ is $\Gamma=\Sigma\cup\{\cents, \$\}$.

A superposition of $M$ is any element of $l_2(Q)$
(the space of mappings from $Q$ to $\bbbc$ with $l_2$ norm).
For $q\in Q$, $|q\rangle$ denotes the unit vector with
value 1 at $q$ and 0 elsewhere. 
All elements of $l_2(Q)$ can be expressed as linear
combinations of vectors $|q\rangle$.
We will use $\psi$ to denote elements of $l_2(Q)$.

The transition function $\delta$ maps $Q\times\Gamma\times Q$ to $\bbbc$.
The value $\delta(q_1, a, q_2)$ is the amplitude of $|q_2\rangle$ in
the superposition of states to which $M$ goes from $|q_1\rangle$ after
reading $a$. For $a\in\Gamma$, $V_a$ is a linear transformation on $l_2(Q)$
defined by
\begin{equation}
\label{E1}
V_a(|q_1\rangle)=\sum_{q_2\in Q}\delta(q_1, a, q_2)|q_2\rangle.
\end{equation}
We require all $V_a$ to be unitary.

%A configuration of $M$ is described by $s\in l_2(Q)\times\bbbr\times\bbbr$ 
%where $s=(\psi, p_{acc}, p_{rej})$, $p_{acc}\geq 0$, $p_{rej}\geq 0$ 
%and $|\psi|\+p_{acc}+p_{rej}=1$. 
%We interpret $p_{acc}$ as the probability that $M$
%has already accepted the input and $p_{rej}$ as 
%the probability that $M$ has rejected the input.
%$\|\psi\|$ is the probability that $M$ has neither accepted
%nor rejected the input so far and $\psi$ is the current superposition
%of internal states. 

The computation of a QFA starts in the superposition
$|q_0\rangle$. Then transformations corresponding to 
the left endmarker $\cents$, the letters of the input word $x$ and
the right endmarker $\$$ are applied. 
The transformation corresponding to $a\in\Gamma$ consists of two steps.
\begin{enumerate}
\item
First, $V_a$ is applied.
The new superposition $\psi'$ is $V_a(\psi)$
where $\psi$ is the superposition before this step.
\item
Then, $\psi'$ is observed with respect to 
the observable $E_{acc}\oplus E_{rej}\oplus E_{non}$
where $E_{acc}=span\{|q\rangle : q\in Q_{acc}\}$,
$E_{rej}=span\{|q\rangle : q\in Q_{rej}\}$,
$E_{non}=span\{|q\rangle : q\in Q_{non}\}$.
This observation gives $x\in E_{i}$ with the probability
equal to the amplitude of the projection of $\psi'$.
After that, the superposition collapses to this projection.

If we get $\psi'\in E_{acc}$, the input is accepted.
If $\psi'\in E_{rej}$, the input is rejected.
If $\psi'\in E_{non}$, the next transformation is applied.
\end{enumerate}
We regard these two transformations as reading a letter $a$.
%$V'_a$ is the transformation that maps $\psi$
%to the non-halting part of $V_a(\psi)$.
%$V'_a=P_{non} V_a$ where $P_{non}(\psi)$ is a linear transformation which 
%leaves all non-halting components of the configuration 
%$\psi$ unchanged and maps all accepting and rejecting components to $0$. 
%If $x$ is a word consisting of letters $a_1\ldots a_k$,
%then $V_x$ denotes $V_{a_k}\ldots V_{a_1}$ and $V'_x$ denotes
%$V'_{a_k}\ldots V_{a_1}$.

%For a word $x$, $\psi_{x}$ is the non-halting part of the QFA's
%configuration after reading $x$. It is easy to see that, for
%any word $x$ and letter $a$, $\psi_{xa}=V'_a(\psi_x)$.

{\bf Another definition of QFAs.}
Independently of \cite{KW}, quantum automata were 
introduced in \cite{CM}. There is one difference between
these two definitions. In \cite{KW}, a QFA is observed after
reading each letter (after doing each $V_a$).
In \cite{CM}, a QFA is observed only after all letters have been
read. %It is easy to show that 
Any language recognized by a QFA
according to the definition of \cite{CM} is recognized by
a QFA according to \cite{KW}. The converse is not true.
Any finite language can be recognized in the sense
of \cite{KW}. However, no finite non-empty language
can be recognized in the sense of \cite{CM}.
Everywhere in this paper, we will use the more general
definition of \cite{KW}. However, our results of section \ref{S41}
which show that 1-way QFAs can be more space-efficient than
deterministic or probabilistic automata are true in the 
more restricted model of \cite{CM} as well.

\subsection{Example}

To explain our notation, we give an example of a 1-way QFA. 
To keep it simple, we use a one letter alphabet 
$\Sigma=\{a\}$. The state space is $Q=\{q_0, q_1, q_{acc}, q_{rej}\}$
with the set of accepting states $Q_{acc}=\{ q_{acc}\}$ and
the set of rejecting states $Q_{rej}=\{q_{rej}\}$.
The starting state is $q_0$.

The transition function can be specified in two ways:
by specifying $\delta$ or by specifying $V_{x}$ for all letters $x\in\Gamma$.
These methods are equivalent: all $V_x$ are determined by $\delta$ and
equation (\ref{E1}).
We shall define the automaton by describing $V_x$.
\[ V_a(|q_0\rangle)=\frac{1}{2} |q_0\rangle+\frac{1}{2} |q_1\rangle+
\frac{1}{\sqrt{2}} |q_{rej}\rangle,\]
\[ V_a(|q_1\rangle)=\frac{1}{2} |q_0\rangle+\frac{1}{2} |q_1\rangle
-\frac{1}{\sqrt{2}}|q_{rej}\rangle,\]
\[ V_{\$}(|q_0\rangle)= |q_{rej}\rangle, V_{\$}(|q_1\rangle)= |q_{acc}\rangle.\]
It can be also defined by describing $\delta$. 
For example, $V_a(|q_0\rangle)=\frac{1}{2} |q_0\rangle+\frac{1}{2} |q_1\rangle+
\frac{1}{\sqrt{2}} |q_{rej}\rangle$ would be 
\[ \delta(q_0, a, q_0)=\frac{1}{2}, \delta(q_0, a, q_1)=\frac{1}{2},\]
\[ \delta(q_0, a, q_{acc})=0, \delta(q_0, a, q_{rej})=\frac{1}{\sqrt{2}}.\]
As we see, this is much longer. For this reason, we will mainly use $V_x$
notation.

There are some transitions that we have not described.
For example, $V_a(q_{acc})$ has not been specified. These 
values are not important and can be arbitrary.
We need them to be such that $V_a$ is unitary but this is not difficult.
As long as all specified $V_a(q_i)$ are orthogonal, the remaining
values can be assigned so that the whole $V_a$ is unitary.
In the sequel, we will often shorten descriptions of QFAs
by leaving out transitions that can be defined arbitrarily.

Next, we show how this automaton works on the word $aa$.
\begin{enumerate}
\item
The automaton starts in $|q_0\rangle$. Then, $V_a$ is applied,
giving $\frac{1}{2} |q_0\rangle+\frac{1}{2} |q_1\rangle+
\frac{1}{\sqrt{2}} |q_{rej}\rangle$.
This is observed. Two outcomes are possible. With probability
$(1/\sqrt{2})^2=1/2$, a rejecting state is observed.
Then, the superposition collapses to $|q_{rej}\rangle$,
the word is rejected and the computation terminates.
Otherwise (with probability $1/2$), a non-halting state is observed and the
superposition collapses to $\frac{1}{2} |q_0\rangle+\frac{1}{2} |q_1\rangle$.
In this case, the computation continues.
\item
A simple computation %(we leave the details out) 
shows that $\frac{1}{2} |q_0\rangle+\frac{1}{2} |q_1\rangle$ is mapped
to itself by $V_a$.
After that, a non-halting state is observed. (There are no
accepting or rejecting states in this superposition.)
\item
Then, the word ends and the transformation $V_{\$}$ corresponding
to the right endmarker $\$$ is done. It maps
the superposition to $\frac{1}{2} |q_{rej}\rangle+\frac{1}{2} |q_{acc}\rangle$.
This is observed. With probability $(1/2)^2=1/4$, the rejecting state $q_{rej}$
is observed. With probability $1/4$, the accepting state $q_{acc}$ is observed.
\end{enumerate}
The total probability of accepting is $1/4$, the probability
of rejecting is $1/2+1/4=3/4$.

\subsection{Reversible automata}
\label{S23}

A 1-way reversible finite automaton (RFA) is a QFA with
$\delta(q_1, a, q_2)\in\{0, 1\}$ for all $q_1, a, q_2$.
Alternatively, RFA can be defined as a deterministic automaton
where, for any $q_2, a$, there is at most one state
$q_1$ such that reading $a$ in $q_1$ leads to $q_2$.
We use the same definitions of acceptance and rejection.
States are partitioned into accepting, rejecting and non-halting states and
a word is accepted (rejected) whenever the RFA
enters an accepting (rejecting) state. After that,
the computation is terminated. Similarly to quantum case,
endmarkers are added to the input word.
The starting state is one, accepting (rejecting) states
can be multiple. This makes our model different from
both \cite{Angluin} (where only one accepting state was allowed)
and \cite{Pin} (where multiple starting states with a non-deterministic
choice between them at the beginning were allowed).
We define our model so because we want it to be as close to our
model of QFAs as possible.

Generally, it's hard to introduce probabilism into finite automata
without losing reversibility. However, there are some types
of probabilistic choices that are consistent with reversibility.
For example, we can choose the starting state probabilistically.
The next example shows that such probabilistic choices increase the power of
an automaton.

{\bf Example.}
Consider the language $L=\{ a^{2n+3}|n\in\bbbn\}$. It cannot
be recognized by a 1-way RFA.
However, there are 3 1-way RFAs such that
each word in the language is accepted by 2 of them and
each word not in the language is rejected by 2 out of 3.
Hence, if we choose one of these three automata
equiprobably, $L$ will be recognized with the probability
of correct answer $2/3$.

Probabilistic choices of this type can be easily done in our
model of QFAs. This may lead to a claim that QFAs 
are more powerful than classical reversible automata because
they can do such probabilistic choices. We wish to avoid such 
situations and to separate probabilistic choices from real quantum effects.

Therefore, we define 1-way finite automata with probabilistic choices
(PRFAs) and compare capabilities of QFAs with them. 
A PRFA is a probabilistic finite automaton such that, 
for any state $q_1$ and any $a\in\Gamma$, there is at most
one state $q_2$ such that the probability of passing from $q_2$ to $q_1$
after reading $a$ is non-zero. %(No constraint about halting states.)
Definitions of acceptance and rejection
are similar to QFAs and RFAs. Now, the probabilistic automaton
from the example above becomes a 1-way PRFA.

\begin{Theorem}
\label{T9}
\begin{enumerate}
\item
If a language is accepted by a 1-way RFA, it is accepted by a 1-way PRFA.
\item
If a language is accepted by a 1-way PRFA, it is accepted by a 1-way QFA with
the same probability of correct answer.
\end{enumerate}
\end{Theorem}

\noindent
{\bf Proof:}
Easy.
$\Box$

In section \ref{S32} we will compare the power of 1-way 
QFAs and PRFAs and show that 1-way quantum automata can actually do more
than just probabilistic choices.

\section{Capabilities of RFAs and QFAs}

\subsection{QFAs with probability of correct answer above 7/9}

We characterize the languages recognized by 1-way QFAs in terms
of their minimal automata.
The minimal automaton of a language $L$ is a 1-way
deterministic finite automaton recognizing it with the smallest
number of states. (Note: the minimal automaton can be non-reversible,
even for some languages $L$ that can be recognized by a 1-way RFA.
The extreme case of this is our Theorem \ref{T11} where the smallest
1-way RFA is exponentially bigger than the minimal nonreversible
automaton.) It is well known\cite{Hopcroft} that 
the minimal automaton is unique and can be effectively
constructed.

\begin{Theorem}
\label{T1}
Let $L$ be a language and $M$ be its minimal automaton.
Assume that there is a word $x$ such that $M$ 
contains states $q_1$, $q_2$ satisfying:
\begin{enumerate}
\item
$q_1\neq q_2$,
\item
If $M$ starts in the state $q_1$ and reads $x$,
it passes to $q_2$, 
\item
If $M$ starts in $q_2$ and reads $x$,
it passes to $q_2$, and
\item
$q_2$ is neither ``all - accepting" state, nor ``all - rejecting" state.
\end{enumerate}
Then $L$ cannot be recognized by a 1-way QFA with 
probability at least $7/9+\epsilon$ for any fixed $\epsilon>0$.
\end{Theorem}

\begin{figure}
\label{fig-dfa}
\begin{center}
\epsfxsize=3in
\hspace{0in}
\epsfbox{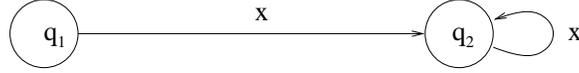}
\caption{\it ``The forbidden construction'' of Theorem 2.}
\end{center}
\end{figure}

\noindent
{\bf Proof:}
We prove the result for a slightly smaller probability of
correct answer $5/6+\epsilon$ (instead of $7/8+\epsilon$).
The proof for $7/8+\epsilon$ is technically more complicated.

Let $L$ be a language such that its minimal automaton
contains the ``forbidden construction" and $M$ be a QFA.
We show that, for some word $y$ the probability of $M$
giving the correct answer to "$y\in L$?" is less than $5/6+\epsilon$.
This implies that $L$ cannot be recognized with probability
of correct answer being $5/6+\epsilon$.

For simplicity, we assume that $q_1$ is the starting state of $M$.
We introduce some notation.
Let $P_{non}(\psi)$ be the non-halting part of $\psi$
and $P_{halt}(\psi)$ be the halting part of $\psi$.
$V'_a=P_{non} V_a$ is a transformation that maps $\psi$
to the non-halting part of $V_a(\psi)$.
%It is easy to see that $V'_a=P_{non} V_a$ where 
%$P_{non}(\psi)$ is a linear tranformation which 
%leaves all non-halting components of the configuration 
%$\psi$ unchanged and maps all accepting and rejecting components to $0$. 
If $x$ is a word consisting of letters $a_1\ldots a_k$,
then %$V_x$ denotes $V_{a_k}\ldots V_{a_1}$ and 
$V'_x$ denotes $V'_{a_k}\ldots V_{a_1}$.
$\psi_{x}$ denotes the non-halting part of the QFA's
configuration after reading $x$. It is easy to see that, for
any word $x$ and letter $a$, $\psi_{xa}=V'_a(\psi_x)$.

We recall that $l_2(Q)$ denotes the state-space of $M$
with $l_2$ norm $\|\psi\|$.
$l_2(Q)=E_{acc}\oplus E_{rej}\oplus E_{non}$.

First, we prove that the state-space of $M$ can be decomposed
into two parts with different behavior.

\begin{Lemma}
\label{LT1}
There are subspaces $E_1$, $E_2$ such that $E_{non}=E_1\oplus E_2$ and
\begin{enumerate}
\item[(i)]
If $\psi\in E_1$, then $V_x(\psi)\in E_1$,
\item[(ii)]
If $\psi\in E_2$, then $\| V'_{x^k}(\psi)\|\rightarrow 0$ when
$k\rightarrow\infty$.
\end{enumerate}
\end{Lemma}

\noindent
{\bf Proof.}
We define two sequences of subspaces $E^1_1, E^2_1, \ldots$ and
$E^1_2, E^2_2, \ldots$ such that $E_{non}=E^i_1\oplus E^i_2$. 
Let $E^1_1=\{\psi|\psi\in E_{non} \mbox{ and } V_a(\psi)\in E_{non}\}$
(i.e., the subspace of all $\psi$ such that both $\psi$ and $V_a(\psi)$
have only non-halting components).
$E^1_2$ consists of all vectors in $E_{non}$ orthogonal to $E^1_1$.
For $i>1$, $E^i_1=E^{i-1}_1\cap \{\psi|V_a(\psi)\in E^{i-1}_1\}$
and $E^i_2$ consists of all vectors in $E_{non}$ orthogonal to $E^i_1$.

Clearly, $E^1_1\supseteq E^2_{1}\supseteq \ldots$.
If $E^{i+1}_1$ is a proper subspace of $E^i_1$,
then the dimensionality of $E^{i+1}_1$ is smaller than the
dimensionality of $E^i_1$. This can happen only finitely many times 
because the original $E^1_1$ is finite-dimensional.
Hence, there is $i_0$ such that $E^{i_0}_1=E^{i_0+1}_1$.
We define $E_1=E^{i_0}_1$, $E_2=E^{i_0}_2$.
Next, we check that both (i) and (ii) are true.

Let $\psi\in E_1$. Then, $V_a(\psi)\in E^{i_0+1}_1=E^{i_0}_1=E_1$
and $V_a(\psi)\in E_{non}$ by $E_1\subseteq E^1_1$ and
the definition of $E_1^1$. It remains to prove that $E_2$ 
satisfies (ii) condition of Lemma \ref{LT1}.

\begin{Claim}
\label{Cl1}
If $\psi\in E^i_1$, then 
$P_{halt} (V_a(V'_{a^l}(\psi)))=\overrightarrow{0}$
for all $l\leq i-1$.
\end{Claim}

\noindent
{\bf Proof:}
By induction.
If $i=1$, then $P_{halt}(V_a(\psi))=\overrightarrow{0}$
by definition of $E^1_1$.
If $i>1$ and $l=0$, then $P_{halt}(V_a(\psi))=\overrightarrow{0}$
because $E^i_1\subseteq E^1_1$.

The only remaining case is $i>1$ and $l>1$.
Let $\psi'=V_a(\psi)$.  By definition of $E^i_1$, $V_a(\psi)\in E^{i-1}_1$. 
We also have $V_a(\psi)\in E_{non}$ because $E^{i-1}_1\subseteq E_{non}$.
Hence, $V'_a(\psi)=P_{non}(V_a(\psi))=V_a(\psi)=\psi'$ and
\[ P_{halt} (V_a(V'_{a^l}(\psi)))=P_{halt}(V_a(V'_{a^{l-1}}(\psi')))=
\overrightarrow{0} \]
by $\psi'\in E^{i-1}_1$ and inductive assumption.
$\Box$

\begin{Claim}
\label{C2}
Let $\psi=\psi_1+\psi_2$, $\psi_1\in E^i_1, \psi_2\in E^i_2$.
Then, for all $l\leq i-1$,
\[ P_{halt} (V_a(V'_{a^l}(\psi)))=P_{halt} (V_a(V'_{a^l}(\psi_2))).\]
\end{Claim}

\noindent
{\bf Proof:}
By linearity of $P_{halt}, V_a, V'_a$,
\[ P_{halt} (V_a(V'_{a^l}(\psi)))=P_{halt} (V_a(V'_{a^l}(\psi_1)))+
P_{halt} (V_a(V'_{a^l}(\psi_2))).\]
Claim \ref{Cl1} implies that 
$P_{halt} (V_a(V'_{a^l}(\psi_1)))=\overrightarrow{0}$.
$\Box$

\begin{Claim}
\label{C3}
Let $j\in\{1, \ldots, i_0\}$.
There is a constant $\delta_j>0$ such that for any $\psi\in E^i_2$
there is $l\in\{0, \ldots, j-1\}$ such that 
$\|P_{halt} (V_a(V'_{a^l}(\psi)))\|\geq \delta_j\|\psi\|$. 
\end{Claim}

\noindent
{\bf Proof:}
By induction.

\noindent
{\bf Base Case.}
Consider the linear transformation 
$T_1:E^1_2\rightarrow E_{acc}\oplus E_{rej}$ that maps $\psi\in E$ to
the halting part of $V_a(\psi)$. $\|T_1\|$ (the norm of $T_1$) is
the minimum of $\|T_1(x)\|$ over all $x$ such that $\|x\|=1$.

If $\|T_1\|=0$, there is $x\in E^1_2$ such that $\|x\|=1$ but 
$\|T_1(x)\|=0$. This means that $T_1(x)=\overrightarrow{0}$, 
implying that $x\in E^1_1$. However, 
$E^1_1\cap E^1_2=\{\overrightarrow{0}\}$, 
leading to a contradiction.
Hence, $\|T_1\|>0$. Also, $\|T_1\|\leq 1$ because
$V_a$ is unitary and projection to the halting subspace
can only decrease the norm.

We take $\delta_1=\|T_1\|$.
Then, the halting part of $V_a(\psi)$ is at least 
$\|T_1\|\|\psi\|=\delta_1\|\psi\|$.

{\bf Inductive Case.}
We assume that the lemma is true for $E^i_2$ and prove it for $E^{i+1}_2$.

We consider the transformation $T_{i+1}$ mapping $\psi\in E^{i+1}_2$
to the projection of $V_a(\psi)$ on $E^i_2$. If 
$T_{i+1}(\psi)=\overrightarrow{0}$, then $\psi\in E^{i+1}_1$ by
the definition of $E^{i+1}_1$. Similarly to the previous case,
$E^{i+1}_1\cap E^{i+1}_2=\{\overrightarrow{0}\}$.
Hence, if $T_{i+1}(\psi)=\overrightarrow{0}$ and $\psi\in E^{i+1}_2$,
then $\psi=\overrightarrow{0}$. This means that $\|T_{i+1}\|>0$.
We can also prove that $\|T_{i+1}\|\leq 1$.

We define $\delta_{i+1}=\frac{\|T_{i+1}\|}{4} \delta_i$.

Let $E^i_3=\{ x\in E^{i+1}_2 \mbox{ and } x\bot E^i_2\}$.
Then, $E^{i+1}_2=E^i_2\oplus E^i_3$.
We also note that $E^i_3$ is a subspace of $E^i_1$.
(This follows from definitions of $E^i_1$ and $E^i_3$.)

To show that one of $P_{halt} (V_a(V'_{a^l}(\psi)))$ is large
enough, we represent $\psi\in E^{i+1}_2$ as $\psi_2+\psi_3$,
$\psi_2\in E^i_2$, $\psi_3\in E^i_3$.
There are two cases:

\begin{enumerate}
\item
$\|\psi_2\|\geq \frac{\|T_{i+1}\|}{4}\|\psi\|$.

Then, 
\[ P_{halt} (V_a(V'_{a^l}(\psi_2)))\geq \delta_i \|\psi_2\|\geq
\delta_i \frac{\|T_{i+1}\|}{4}\|\psi\|=\delta_{i+1}\|\psi\|\]
for some $l\leq i-1$ because $\psi_2\in E^i_2$ and we can apply the inductive
assumption. Claim \ref{C2} implies that this is also true
for $P_{halt} (V_a(V'_{a^l}(\psi)))$.

\item
$\|\psi_2\|< \frac{\|T_{i+1}\|}{4}\|\psi\|$.

Then, by triangle inequality,
\[ \|\psi_3\|\geq \|\psi\|-\|\psi_2\|
\geq (1-\frac{\|T_{i+1}\|}{4})\|\psi\|\geq 
\frac{3\|T_{i+1}\|}{4}\|\psi\|.\]

Let $\psi'$, $\psi'_2$ and $\psi'_3$ be the projections 
of $V_a(\psi)$, $V_a(\psi_2)$, $V_a(\psi_3)$ on $E^i_2$.
Clearly, $\psi'=\psi'_2+\psi'_3$.
Triangle inequality gives us
\[ \|\psi'\|\geq \|\psi'_3\|-\|\psi'_2\| \geq \|\psi'_3\|-\|\psi_2\|\geq 
\frac{3\|T_{i+1}\|}{4}\|\psi\|-\frac{\|T_{i+1}\|}{4}\|\psi\|=
\frac{\|T_{i+1}\|}{2}\|\psi\|.\]
We have $\psi'=P_{non}(\psi')+P_{halt}(\psi')$.
Again, we have two cases.

If $\|P_{halt}(\psi')\|\geq \frac{\|T_{i+1}\|}{4}\|\psi\|$,
then $\|P_{halt}(\psi')\|\geq \delta_{i+1}\|\psi\|$ because
$\delta_{i+1}=\frac{\|T_{i+1}\|}{4}\delta_i$ and $\delta_i\leq 1$
because all $\|T_i\|$ are at most 1.

Otherwise, by triangle inequality, 
$\|P_{non}(\psi')\|\geq \|\psi'\|-\|P_{halt}(\psi')\|\geq
\frac{\|T_{i+1}\|}{4}\|\psi\|$.
By inductive assumption, there is $l\leq i-1$ such that
$\|P_{halt} V_a(V'_{a^l}(\psi'))\|\geq \delta_i \|\psi'\|$.
Therefore, 
\[ \|P_{halt} (V_a(V'_{a^{l+1}}(\psi)))\|\geq 
\frac{\|T_{i+1}\|}{4}\delta_i\|\psi\|=\delta_{i+1}\|\psi\|. \]
\end{enumerate}
$\Box$

\begin{Claim}
\label{C4}
There is $c$ such that $0<c<1$ and, for any $\psi\in E_2$, $t\in\bbbn$,
$\|V'_{a^{i_0 t}}(\psi)\|\leq c^t \|\psi\|$.
\end{Claim}

\noindent
{\bf Proof:}
We take $c=\sqrt{1-\delta_{i_0}^2}$.

By Claim \ref{C3}, one of $P_{halt}(V_a(V'_{a^l}(\psi)))$ 
is at least $\delta_{i_0}\|\psi\|$.
$P_{non}(V_a(V'_{a^l}(\psi)))$ is orthogonal to this vector.
Hence, $P_{non}(V_a(V'_{a^l}(\psi)))$ is at most
\[ \sqrt{\|\psi\|^2-\delta^2_{i_0}\|\psi\|^2}=
\sqrt{1-\delta^2_{i_0}}\|\psi\|. \]
$V'_{a^{i_0}}$ can be only smaller because
$V_a$ is unitary and $P_{non}$ does not increase vectors.

We have shown that $V'_{a^{i_0}}(\psi)\leq c \|\psi\|$.
Repeating this $t$ times, we obtain Claim \ref{C4}.
$\Box$

Clearly, $c^t\rightarrow 0$ if $t\rightarrow\infty$.
This completes the proof of Lemma \ref{LT1}.
$\Box$

Let $\psi_{\cents}=\psi^1_{\cents}+\psi^2_{\cents}$,
$\psi^1_{\cents}\in E_1$, $\psi^2_{\cents}\in E_2$.
We consider two cases.

\noindent
{\bf Case 1.}
$\|\psi^2_{\cents}\|>1/\sqrt{3}$.

Then, $\|\psi^1_{\cents}\|<\sqrt{1-(1/\sqrt{3})^2}=2/\sqrt{3}$.
This also means $\|V'_{x^i}(\psi^1_{\cents})\|<2/\sqrt{3}$.
For sufficiently large $i$, $\|V'_{x^i}(\psi^2_{\cents})\|$
becomes negligible (part (ii) of Lemma \ref{LT1}). 
Then, $\|\psi_{\cents x^i}\|<2/\sqrt{3}$.
The probability of $M$ halting after this moment is less than 2/3.
Hence, $M$ has already halted with probability more than 1/3 and 
accepted (or rejected) with probability more than 1/6.
This means that $M$ cannot reject (accept) any continuation of $x^i$
with probability 5/6. However, $x^i$ has both 
continuations in $L$ and continuations not in $L$.
Hence, $M$ does not recognize $L$.

\noindent
{\bf Case 2.}
$\|\psi^2_{\cents}\|\leq 1/\sqrt{3}$.

$q_1$ and $q_2$ are different states in the minimal automaton of $L$.
Therefore, there is a word $y\in\Sigma^{*}$ such that $y$ leads
to acceptance from one of $q_1, q_2$ but not from the other one.
We consider the distributions of probabilities on $M$'s answers ``accept"
and ``reject" on $y$ and $x^i y$.
On one of these words, $M$ must accept with probability at least $5/6+\epsilon$
and reject with probability at most $1/6-\epsilon$.
On the other word, $M$ must accept with probability most $1/6-\epsilon$
and reject with probability at least $5/6+\epsilon$.
Therefore, both the probabilities of accepting and the probabilities of
rejecting must differ by at least $2/3+2\epsilon$.
This means that the {\em variational distance} between two probability
distributions (the sum of these two distances) must be
at least $4/3+4\epsilon$.
We show that it cannot be so large.

First, we select an appropriate $i$.
Let $m$ be so large that $\|V'_{x^m}(\psi^2_{\cents})\|\leq \delta$ for
$\delta=\epsilon/4$.
$\psi^1_{\cents}, V'_x(\psi^1_{\cents}), V'_{x^2}(\psi^1_{\cents})$, $\ldots$
is a sequence in a finite-dimensional space. Therefore, it
has a limit point and there are $i, j$ such that
\[  \|V'_{x^j}(\psi^1_{\cents})-V'_{x^{i+j}}(\psi^1_{\cents})\|<\delta.\]
We choose $i, j$ so that $i>m$.

The difference between two probability distributions comes from two sources.
The first is difference between $\psi_{\cents}$ and
$\psi_{\cents x^i}$ (the states of $M$ before reading $y$).
The second source is the possibility of $M$ accepting while
reading $x^i$ (the only part that is different in two words).
We bound the difference created by each of these two sources.

The difference $\psi_{\cents}-\psi_{\cents x^i}$
can be partitioned into three parts.
\[ \psi_{\cents}-\psi_{\cents x^i}=(\psi_{\cents}-\psi^1_{\cents})+
(\psi^1_{\cents}-\psi^1_{\cents x^i})+
(\psi^1_{\cents x^i}-\psi_{\cents x^i}).\]

The first part is
$\psi_{\cents}-\psi^1_{\cents}=\psi^2_{\cents}$ and
$\|\psi^2_{\cents}\|\leq\frac{1}{\sqrt{3}}$.
The second and the third parts are both small.
For the second part, notice that $V'_x$ is unitary
on $E_1$ (because $V_x$ is unitary and $V_x(\psi)$ does not
contain halting components for $\psi\in E_1$).
Hence, $V'_x$ preserves distances on $E_1$ and
\[ \|\psi^1_{\cents}-\psi^1_{\cents x^i}\|=
\|\psi^1_{\cents x^j}-\psi^1_{\cents x^{i+j}} \|\leq\delta.\]

The third part is $\psi_{\cents x^i}-\psi^1_{\cents x^i}=
\psi^2_{\cents x^i}$ and $\|\psi^2_{\cents x^i}\|\leq \delta$ because
$i>m$.

Next, we state two lemmas relating differences
between two superpositions and differences between probability
distributions created by observing these superpositions.
The first lemma is by Bernstein and Vazirani\cite{BV}.

\begin{Lemma}
\cite{BV}
\label{BVTheorem}
Let $\psi$ and $\phi$ be such that $\|\psi\|\leq 1$, $\|\phi\|\leq 1$ and
$\|\psi-\phi\|\leq\epsilon$.
Then the total variational distance resulting from measurements
of $\phi$ and $\psi$ is at most $4\epsilon$.
\end{Lemma}

The second lemma is our improvement of lemma \ref{BVTheorem}.

\begin{Lemma}
\label{BVPlus}
Let $\psi^1$ and $\psi^2$ be such that $\psi^1\bot\psi^2$.
Then the total variational distance resulting from measurements
of $\psi^1$ and $\psi^1+\psi^2$ is at most
\[ \|\psi^2\|\sqrt{\|4\psi^1\|^2+\|\psi^2\|^2}.\]
\end{Lemma}

\noindent
{\bf Proof.}
Omitted.
$\Box$

We apply lemma \ref{BVPlus} to 
$\psi^1_{\cents}$ and $\psi^2_{\cents}$.
This gives that the variational distance between
distributions generated by $\psi^1_{\cents}$ and
$\psi^1_{\cents}+\psi^2_{\cents}$ is at most 1.
Then, we apply lemma \ref{BVTheorem} to two other parts
of $\psi_{\cents}-\psi_{\cents x^i}$.
Each of them influences the variational distance by at most $4\delta$.
Together, the variational distance between distributions obtained
by observing $\psi_{\cents}$ and $\psi_{\cents x^i}$ is at most
$1+8\delta$.

The probability of $M$ halting while reading $x^i$ is at most
$\|\psi^2_{\cents}\|^2=1/3$. Adding it increases the
variational distance by at most 1/3.
Hence, the total variational distance is at most $4/3+8\delta=
4/3+2\epsilon$.
However, if $M$ distinguishes $y$ and $x^i y$ correctly,
it should be at least $4/3+4\epsilon$.
Hence, $M$ does not recognize one of these words correctly.
$\Box$

\begin{Theorem}
\label{T2}
Let $L$ be a language and $M$ be its minimal automaton.
If $M$ does not contain the ``forbidden construction'' of
Theorem \ref{T1},
%If there is no $q_1, q_2, x$ such that
%\begin{enumerate}
%\item
%$q_1\neq q_2$,
%\item
%If $M$ starts in the state $q_1$ and reads $x$,
%it passes to $q_2$, 
%\item
%If $M$ starts in the state $q_2$ and reads $x$,
%it passes to $q_2$, and
%\item
%$q_2$ is neither "all-accepting" state, nor "all-rejecting" state,
%\end{enumerate}
then $L$ can be recognized by a 1-way reversible finite automaton.
\end{Theorem}

\noindent
{\bf Proof.}
We define a non-reversibility as a tuple 
$\langle q_1, q_2, q, a\rangle$ where $q_1, q_2, q\in Q$,
$a\in\Sigma$, $q_1\neq q_2$ and reading $a$ in $q_1$ or $q_2$
leads to $q$. Let $m$ be the number of non-reversibilities 
in $M$. We show how to modify $M$ so that the number of
non-reversibilities decreases. A reversible automaton is
obtained by repeating this modification several times.

We define a partial ordering $<$ on non-reversibilities.
$\langle q_1, q_2, q, a\rangle<\langle q'_1, q'_2, q', a'\rangle$
if and only if one of $q'_1$ and $q'_2$ is reachable
from $q$. It is easy to see that $<$ is transitive.

\begin{Lemma}
$<$ is anti-reflexive.
\end{Lemma}

\noindent
{\bf Proof.}
For a contradiction, assume there is $\langle q_1, q_2, q, a\rangle$
such that $\langle q_1, q_2, q, a\rangle<\langle q_1, q_2, q, a\rangle$.
We also assume that $q_2$ is reachable from $q$ by reading a word $y$. 
(Otherwise, $q_1$ is reachable from $q$ and we can just exchange $q_1$
and $q_2$.) Then, reading $x=ay$ leads from $q_1$ to $q_2$ and
from $q_2$ to $q_2$. This contradicts our assumption that $M$
does not contain such $q_1, q_2$.
$\Box$

Hence, there is a tuple $\langle q_1, q_2, q, a\rangle$ that is
maximal with respect to $<$. We create two copies for
state $q$ and all states reachable from $q$. 
If $M$ reads $a$ in $q_1$, it passes to one copy of
$q$, if it reads $a$ in $q_2$, it passes to the second copy.
This eliminates this non-reversibility. 
Other non-reversibilities are not dublicated because
they are not reachable from $q$. 
Hence, the number of non-reversibilities is decreased.
$\Box$

\begin{Corollary}
A language can be recognized by a 1-way QFA with probability 
$7/9+\epsilon$ if and only if it can be recognized by a
1-way reversible finite automaton.
\end{Corollary}

\noindent
{\bf Proof.}
Clearly, a RFA is a special case of a QFA.
The other direction follows from Theorems \ref{T1} and \ref{T2}.
$\Box$ 

This immediately implies the same result about 1-way reversible 
automata with probabilistic choices. 
For this type of automata, a stronger result can be proved.

\begin{Theorem}
\label{T3}
A language can be recognized by a 1-way PRFA with probability 
$2/3+\epsilon$ (for arbitrary $\epsilon>0$)
if and only if it can be recognized by a
1-way reversible finite automaton.
\end{Theorem}

\noindent
{\bf Proof:}
Omitted.
$\Box$

The example in Section \ref{S23} shows that Theorem \ref{T3} is tight.
%Probably, the constant $7/9$ in Theorem \ref{T1} can be slightly improved
%by doing the counting more precisely.

\subsection{QFAs with probability of correct answer below 7/9}
\label{S32}

For smaller probabilities, QFAs are slightly more powerful than
RFAs or even PRFAs. 

\begin{Theorem}
\label{T4}
The language $a^{*}b^{*}$ can be recognized by a 1-way QFA
with the probability of correct answer $p=0.68...$ where
$p$ is the root of $p^3+p=1$.
\end{Theorem}

\noindent
{\bf Proof.}
We describe a 1-way QFA $M$ accepting this language. 
The automaton has 4 states: $ q_0,  q_1,  q_{acc}$ and $ q_{rej}$.
$Q_{acc}=\{ q_{acc}\}$, $Q_{rej}=\{ q_{rej}\}$.
The initial state is $\sqrt{1-p} |q_0\rangle+\sqrt{p} |q_1\rangle$.
The transition function is 
\[ V_a(|q_0\rangle)=(1-p) |q_0\rangle+\sqrt{p(1-p)} |q_1\rangle+
\sqrt{p} |q_{rej}\rangle,\]
\[ V_a(|q_1\rangle)=\sqrt{p(1-p)} |q_0\rangle+p |q_1\rangle-\sqrt{1-p}
|q_{rej}\rangle,\]
\[ V_b(|q_0\rangle)= |q_{rej}\rangle, V_b(|q_1\rangle)= |q_1\rangle,\]
\[ V_{\$}(|q_0\rangle)= |q_{rej}\rangle, V_{\$}(|q_1\rangle)= |q_{acc}\rangle.\]

\noindent
{\em Case 1.}
The input is $x=a^*$.

It is straightforward that $\delta$ maps 
$\sqrt{1-p}|q_0\rangle+\sqrt{p}|q_1\rangle$ to itself while it receives $a$
from the input. Hence, after reading $a^*$ the state remains
$\sqrt{1-p}|q_0\rangle+\sqrt{p}|q_1\rangle$
and, after reading the right endmarker,
it becomes $\sqrt{1-p}|q_{rej}\rangle+\sqrt{p}|q_{acc}\rangle$.
This means that the automaton accepts with probability $p$.

\noindent
{\em Case 2.}
The input is $x=a^*b^+$.

Again, the state remains $\sqrt{1-p}|q_0\rangle+\sqrt{p}|q_1\rangle$
while input contains $a$. Reading the first $b$ changes it
to $\sqrt{1-p}|q_{rej}\rangle+\sqrt{p}|q_1\rangle$.
The non-halting part of this state
is $\sqrt{p}|q_1\rangle$. It is left unchanged by next $b$s and mapped
to $|q_{acc}\rangle$ after reading the right endmarker.
Again, the accepting probability is $p$.

\noindent
{\em Case 3.}
The input is $x\notin a^*b^*$.

Then, the initial segment of $x$ is $a^*b^+a^+$.
After reading the first $b$, the state is
$\sqrt{1-p}|q_{rej}\rangle+\sqrt{p}|q_1\rangle$.
The automaton rejects at this moment with probability $(1-p)$.
The non-halting part $\sqrt{p}|q_1\rangle$ is mapped to 
$p\sqrt{1-p}|q_0\rangle+(1-p)\sqrt{p} |q_1\rangle
-\sqrt{p(1-p)}|q_{rej}\rangle$ by the next $V_a$.
Then, the automaton rejects with probability $p(1-p)$.
The non-halting part $p\sqrt{1-p} |q_0\rangle+(1-p)\sqrt{p} |q_1\rangle$
is unchanged by $a$s. However, either $b$ or right endmarker
follows $a$s and then $ q_0$ is mapped to $|q_{rej}\rangle$ and the automaton
rejects with probability $p^2(1-p)$.
We add the probabilities of rejecting at different moments together
and get that $M$ rejects $x\notin a^* b^*$ with probability at least
\[ (1-p)+p(1-p)+p^2(1-p)=(1+p+p^2)(1-p)=\]
\[ \frac{1-p^3}{1-p}(1-p)=1-p^3=p.\]
$\Box$

It is easy to see that the minimal automaton of $a^*b^*$ contains
the ``forbidden construction'' of Theorem \ref{T1}.
Therefore, we have

\begin{Corollary}
\label{C1}
There is a language that can be recognized by a 1-QFA with probability
$0.68...$ but not with probability $7/9+\epsilon$.
\end{Corollary}

\noindent
{\bf Proof:}
Follows from Theorems \ref{T1} and \ref{T4}.
$\Box$

For probabilistic computation, the property that the probability 
of correct answer can be increased arbitrarily is considered evident.
Hence, it was not surprising that \cite{KW} wrote
``with error probability bounded away from $1/2$" about QFAs, 
thinking that all such probabilities are equivalent.
However, mixing reversible (quantum computation) and nonreversible
(measurements after each step) components in one model
makes it impossible for QFAs.
It is open whether a counterpart of Corollary \ref{C1} is true
for 2-way QFAs.    

\begin{Corollary}
There is a language that can be recognized by a 1-QFA with probability
$0.68...$ but not by a classical 1-way reversible FA.
\end{Corollary}

This corollary can be improved by showing that even
a 1-way probabilistic reversible automaton cannot recognize 
this language (and even with probability $1/2+\epsilon$).
%(Example in section \ref{S2} shows that there are some other languages that can be
%recognized by a probabilistic reversible automaton but
%not by reversible automata.) 

\begin{Theorem}
\label{T5}
Let $L$ be a language and $M$ be its minimal automaton. 
Assume that there are words $x, y$ and $M$'s states $q_1, q_2$
such that 
\begin{enumerate}
\item 
none of $q_1$ and $q_2$ is "all-accepting" or "all-rejecting" state;
\item
reading $x$ in $q_1$ leads to $q_1$;
\item
reading $y$ in $q_1$ leads to $q_2$;
\item
reading $y$ in $q_2$ leads to $q_2$;
\item
there is no $i>0$ such that reading $x^i$ leads from $q_2$ to $q_2$.
\end{enumerate}
Then $L$ cannot be recognized by a 1-way PRFA with probability
$1/2+\epsilon$, for any $\epsilon>0$.
\end{Theorem}

\noindent
{\bf Proof.}
Without the loss of generality, we assume that $q_1$ is
the starting state of $M$. 
Let $M_p$ be a 1-way probabilistic reversible automaton.
We are going to show that, for some word $x$, the probability of
$M_p$ giving the right answer on the input $x$ is less than $1/2+\epsilon$.
%We will show that there are $i, j, k>0$ such that
%the difference between the state of $M_p$ after reading 
%$x^i y^j$ and the state after reading $x^i y^j x^k$
%is arbitrarily small. 

\begin{Lemma}
\label{L1}
For any state $q$ and $a\in\Sigma^{+}$, there is $k$ such that
$0<k\leq |Q|$ and, for any sequence of probabilistic choices, 
one of the following happens:
\begin{enumerate}
\item
After reading $a^k$ in state $q$, $M_p$ returns to $q$;
\item
After reading $a^{|Q|+1}$ in state $q$, $M_p$ accepts or rejects.
\end{enumerate}
\end{Lemma}

\noindent
{\bf Proof.}
Let $q_0, q_1, \ldots$ be any sequence of 
non-halting states such that $q_0=q$ and
the probability that reading $a$ causes $M_p$ to go from $q_i$
to $q_{i+1}$ is non-zero. If the length of the sequence is greater than 
$|Q|$, then some state appears twice in this sequence.
We consider the first state which appears twice. If it is not
$q_0$, then it has two preceding states: the state preceding it when it
appears in the sequence for the first time and the state preceding it
when it appears in the sequence for the second time. This contradicts
the definition of a probabilistic reversible automaton.
We have shown that $q_0$ is the first state which appears twice.

Next, assume we have two such sequences: $q_0, q_1, \ldots$ and
$q'_0, q'_1, \ldots$. Let $k_1$, $k_2$ be the smallest numbers such that
$k_1>0$, $q_{k_1}=q_0$ and $k_2>0$, $q'_{k_2}=q'_0$, respectively.
We show that $k_1=k_2$. For a contradiction, assume that $k_1>k_2$
($k_2>k_1$ case is similar.). Then, $q_{k_1-1}=q'_{k_2-1}$ 
(because the state $q_{k_1}=q'_{k_2}$ cannot
have two preceding states), $q_{k_1-2}=q'_{k_2-2}$ and so on,
$q_{k_1-k_2}=q'_{k_2-k_2}=q'_0=q_0$.
This contradicts the assumption that $k_1$ is the smallest
number such that $k_1>0$ and $q_{k_1}=q_0$.
Hence, $k_1=k_2$.
$\Box$

\begin{Lemma}
\label{L2}
For any state $q$ and $a\in\Sigma^{+}$, one of the following happens:
\begin{enumerate}
\item
There is $k$ such that, after reading $a^k$ in state $q$, $M_p$ returns to $q$
for any sequence of probabilistic choices;
\item
The probability of halting after reading $a^k$ in state $q$ tends to 1
when $k\rightarrow\infty$.
\end{enumerate}
\end{Lemma}

\noindent
{\bf Proof.}
Let $k$ be as in Lemma \ref{L1}. If $M_p$ always returns to $q$ after 
reading $a^k$, Lemma \ref{L2} is true. 
It remains to consider the case if there is a sequence of probabilistic
choices for which $M_p$ does not return to $q$.
Then, by Lemma \ref{L1}, this sequence causes $M_p$ to halt.
Let $p$ be the probability of returning to $q$ after reading $a^k$.
Then, the probability of returning to $q$ after reading $a^{ik}$ is
$p^i$. With probability $1-p^i$, $M_p$ does not returns to $q$ at some
moment and (this is the only alternative) terminates after reading $a^{ik+|Q|}$
(or some its prefix). Clearly, $1-p^i\rightarrow 1$, if $i\rightarrow\infty$.
$\Box$

We note that one can use the same $k=|Q|!$ for all $q$ and $a$.
(For any $q, a$, $k\leq |Q|$ and $|Q|!$ is a multiple of any such $k$.)
We shall call the states of the first type {\em return states} for $a$.

Let $p_i$ be the probability of non-halting after reading $x^i$
and $p=\lim_{i\rightarrow\infty}p_i$. 
We select $i$ so that $|p-p_i|<\epsilon$.
Let $p'_j$ be the probability of non-halting after reading $x^i y^j$
and $p'=\lim_{j\rightarrow\infty}p'_j$.
We select $j$ so that $j$ is a multiple of $|Q|!$ and 
$|p'-p'_j|<\epsilon$.
 
Next, we compare the behaviour of $M_p$ on $x^i y^j$ and $x^i y^j x^{|Q|!}$.
These words correspond to different states in the minimal automaton.
Hence, there is a continuation $z$ such that exactly one of 
$x^i y^j z$ and $x^i y^j x^{|Q|!}z$ is in $L$.

If $M_p$ had accepted or rejected after $x^i y^j$ (without seeing the right 
endmarker), it accepts (rejects) both $x^i y^j z$ and $x^i y^j x^{|Q|!}z$.
It remains to consider the sequences of probabilistic choices where 
$M_p$ does not accept until $x^i y^j$.

Let $q_x$ be the state of $M_p$ after reading $x^i$. 
We consider three cases:
\begin{enumerate}
\item
$q_x$ is not a return state for $x$.

Then, reading more $x$'s cause $M_p$ to halt with probability 1.
However, the probability of halting after reading more than $i$ $x$'s
is less than $\epsilon$ (by the definition of $i$).
Hence, the probability of this case is less than $\epsilon$.
\item
$q_x$ is not a return state for $y$.

Then, reading $y$'s cause $M_p$ to halt with probability 1.
If it does not happen before reading $y^j$, it happens later
with probability 1. The definition of $j$ implies that,
if $M$ does not halt before reading $y^j$, then the probability of it
halting later is less than $\epsilon$. Hence, the probability that $q_x$
is not a return state and $M_p$ does not halt before reading $x^i y^j$ is
less than $\epsilon$.
\item
$q_x$ is return state for both $x$ and $y$.

Then, reading $y^{j}$ causes $M_p$ to return to $q_x$ because $j$ is a 
multiple of $|Q|!$ and reading $x^{|Q|}$ causes it to return to $q_x$ as well.
In both cases, it is in the same state after reading $x^i y^j$  and
after reading $x^i y^j x^{|Q|}$ and, hence, does the same
thing on both  $x^i y^j z$ and $x^i y^j x^{|Q|!}z$.
\end{enumerate}
We see that the third case causes $M_p$ to react similarly on 
$x^i y^j z$ and $x^i y^j x^{|Q|!}z$ and the probability of the other
two cases together is less than $2\epsilon$.
Hence, the probabilities of accepting these two words differ by less
than $2\epsilon$. However, one of them is in $L$ and 
must be accepted with probability $1/2+\epsilon$ and the second
is not in $L$ and must be accepted with probability at most $1/2-\epsilon$.
This means that $M_p$ does not recognize $L$ with probability $1/2+\epsilon$.
$\Box$

The ``forbidden construction'' of Theorem \ref{T5} is also 
present in the minimal automaton of $a^*b^*$.
Therefore, we have

\begin{Corollary}
There is a language that can be recognized by a 1-QFA with probability
$0.68...$ but cannot be recognized by a 1-PRFA with probability
$1/2+\epsilon$, for any $\epsilon>0$.
\end{Corollary}

%\noindent
%{\bf Proof:}
%Follows from Theorems \ref{T4} and \ref{T5}.
%$\Box$

We do not know whether all languages with minimal automata not
containing the construction in Theorem \ref{T5} can 
be recognized by 1-way PRFAs.
Another open question is characterizating
the languages recognized by 1-way QFAs
in terms of ``forbidden constructions".

\section{Complexity}
\label{S4}

\subsection{Divisibility by a prime}
\label{S41}

All previous work on 1-way QFAs (\cite{KW,CM} and the previous
sections of this paper) considers %the capabilities of such automata,
the question what languages can be recognized by quantum automata.
However, there is another interesting and important question:
how efficient are QFAs compared to their classical 
counterparts?

For 1-way finite automata, the most natural complexity measure is
the number of states in the automaton. 
We can follow the proof in \cite{KW} that any language recognized by a 1-way
QFA is regular step by step and add complexity bounds to it. Then, we get

\begin{Theorem}
Let $L$ be a language recognized by a 1-way QFA
with $n$ states. Then it can be recognized by a 1-way deterministic 
automaton with $2^{O(n)}$ states. %where $a$ is a constant depending on the
%probability with which $L$ gives correct answer.
\end{Theorem}

So, transforming a QFA into a classical automaton
can cause an exponential increase in its size. Our next results show
that, indeed, 1-way QFAs can be exponentially smaller than their 
classical counterparts.

Let $p$ be a prime. We consider the language  
$L_p=\{ a^i |\mbox{$i$ is divisible by $p$}\}$.
It is easy to see that any deterministic 1-way finite 
automaton recognizing $L_p$ has at least $p$ states.
However, there is a much more efficient QFA!

%\begin{Theorem}
%\label{T7}
%Any deterministic 1-way finite automaton recognizing $L_p$ has
%at least $p$ states.
%\end{Theorem}

%\noindent
%{\bf Proof.}
%Easy.
%$\Box$

\begin{Theorem}
\label{T6}
For any $\epsilon>0$, there is a QFA with $O(\log p)$ states recognizing
$L_p$ with probability $1-\epsilon$.
\end{Theorem}

\noindent
{\bf Proof.}
First, we construct an automaton accepting all words in $L$ with probability 
1 and accepting words not in $L$ with probability at most $7/8$.
Later, we will show how to increase the probability of correct answer
to $1-\epsilon$ for an arbitrary constant $\epsilon>0$.

Let $U_k$, for $k\in\{1, \ldots, p-1\}$ be a quantum automaton 
with a set of states $|Q|=\{q_0, q_1, q_{acc}, q_{rej}\}$, 
a starting state $|q_0\rangle$,
$Q_{acc}=\{q_{acc}\}$, $Q_{rej}=\{ q_{rej}\}$.
The transition function is defined as follows.
Reading $a$ maps $|q_0\rangle$ to $\cos\phi|q_0\rangle+i \sin\phi|q_1\rangle$
and $|q_1\rangle$ to $i \sin\phi|q_0\rangle+\cos\phi|q_1\rangle$
where $\phi=\frac{2\pi k}{p}$.
(It is easy to check that this transformation is unitary.)
Reading the right endmarker $\$$ maps $|q_0\rangle$ to $|q_{acc}\rangle$ and
$|q_1\rangle$ to $|q_{rej}\rangle$.

\begin{Lemma}
\label{L3}
After reading $a^j$, the state of $U_k$ is
\[ \cos\left(\frac{2\pi j k}{p}\right)|q_0\rangle
+i\sin\left(\frac{2\pi j k}{p}\right)|q_1\rangle.\]
\end{Lemma}

\noindent
{\bf Proof.}
By induction. 
$\Box$

If $j$ is divisible by $p$, then $\frac{2\pi j k}{p}$
is a multiple of $2\pi$, $\cos(\frac{2\pi j k}{p})=1$,
$\sin(\frac{2\pi j k}{p})=0$, reading $a^j$ maps
the starting state $|q_0\rangle$ to $|q_0\rangle$ and 
the right endmarker $\$$ maps it to $|q_{acc}\rangle$.
Therefore, all automata $U_k$ accept words in $L$ with probability 1.

For a word $a^j\notin L$, call $U_k$ ``good'' 
if $U_k$ rejects $a^j$ with probability at least 1/2.

\begin{Lemma}
\label{L4}
For any $a^j\notin L$, at least $(p-1)/2$ of all $U_k$ are ``good''.
\end{Lemma}

\noindent
{\bf Proof.}
The superposition of $U_k$ after reading $a^j$ is 
$\cos(\frac{2\pi j k}{p})|q_0\rangle+\sin(\frac{2\pi j k}{p})|q_1\rangle$.
This is mapped to $\cos(\frac{2\pi j k}{p})|q_{acc}\rangle
+\sin(\frac{2\pi j k}{p})|q_{rej}\rangle$ by the right endmarker.
Therefore, the probability of $U_k$ accepting $a^i$ 
is $\cos^2(\frac{2\pi j k}{p})$. 
$\cos^2(\frac{2\pi j k}{p})\leq 1/2$ if and only if
$|\cos(\frac{2\pi j k}{p})|\leq 1/\sqrt{2}$.
This happens if and only if $\frac{2\pi j k}{p}$ is in 
$[2\pi l+\pi/4, 2\pi l+3\pi/4]$ or in $[2\pi l+5\pi/4, 2\pi l+7\pi/4]$
for some $l\in\bbbn$. 

%Next, we show that there are at least $(p-1)/2$ such $k$.
$\frac{2\pi(jk \bmod p)}{p}\in[\pi/4, 3\pi/4]$ if and only if
$\frac{2\pi jk}{p}\in[2\pi l+\pi/4, 2\pi l+3\pi/4]$ for some $l$.
%Therefore, we can consider $\frac{2\pi (jk\bmod p)}{p}$'s instead of
%$\frac{2\pi jk}{p}$'s.
$p$ is a prime and $j$ is relatively prime with $p$.
Therefore, $j\bmod p$, $2j\bmod p$, $\ldots$, $(p-1)j\bmod p$
are just $1$, $2$, $\ldots$, $p-1$ in different order.
Hence, it is enough to count $k$ such that 
$\frac{2\pi k}{p}\in[\pi/4, 3\pi/4]$ or $\frac{2\pi k}{p}\in[5\pi/4, 7\pi/4]$.

We do the counting for $p=8m+1$. (Other cases are similar.)
Then $\frac{2\pi k}{p}\in[\pi/4, 3\pi/4]$ if and only if 
$m+1\leq k\leq 3m$ and $\frac{2\pi k}{p}\in[5\pi/4, 7\pi/4]$
if and only if $5m+1\leq k\leq 7m$. 
Together, this gives us $4m=(p-1)/2$ ``good" k's.
$\Box$

Next, we consider sequences of $\lceil 8\ln p\rceil$ $k$'s.
A sequence is {\em good} for $a^j$ if at least $1/4$ of all its elements are
good for $a^j$.

\begin{Lemma}
\label{L5}
There is a sequence of length $\lceil 8\ln p\rceil$ which is
good for all $a^j\notin L$.
\end{Lemma}

\noindent
{\bf Proof.}
First, we show that at most $1/p$ fraction of all sequences is not good
for any fixed $a^j\notin L$.

We select a sequence randomly by selecting each of its elements uniformly
at random from $\{1, \ldots, p-1\}$. 
The probability of selecting a good $k$ in each step is at least $1/2$. 
By Chernoff bounds, the probability that less than
$1/4=1/2-1/4$ fraction of all elements is good is at most 
\[ e^{-2 (1/4)^2 8\ln p}=\frac{1}{p}. \]

Hence, the fraction of sequences which are bad for at least one 
$j\in\{1, 2, \ldots, p-1\}$ is at most $(p-1)/p$ and there
is a sequence which is good for all $j\in\{1, \ldots, p-1\}$.
This sequence is good for $a^j\notin L$ with $j>p$ as well because
any $U_k$ returns to the starting state after reading $a^p$ and,
hence, works in the same way on $a^j$ and $a^{j \bmod p}$. 
$\Box$

Next, we use a good sequence $k_1, \ldots, k_{\lceil 8\ln p\rceil}$
to construct a quantum automaton recognizing $L_p$.
The automaton consists of $U_{k_1}$, $U_{k_2}$, $\ldots$, 
$U_{k_{\lceil 8\ln p\rceil}}$ and a distinguished starting state.
Upon reading the left endmarker $\cents$, it passes from the starting state
to a superposition where $|q_0\rangle$ states of all $U_{k_l}$ have 
equal amplitudes. 

Words in $L$ are always accepted because all $U_k$ accept them.
For any $a^j\notin L$, at least $1/4$ of the sequence is good.
This means that at least $1/4$ of all $U_{k_l}$ reject it with probability 
at least $1/2$ and the total probability of rejecting any $a^j\notin L$ is
at least $1/8$. 

Finally, we sketch how to increase the probability of correct answer
to $1-\epsilon$ for an arbitrary $\epsilon>0$.
We do it by increasing the probability of correct answer
for each $U_k$.

Namely, we consider an automaton $U'_k$ with $2^d$ non-halting states 
where $d$ is a constant depending on the required probability $1-\epsilon$.
The states are labelled by strings of 0s and 1s of length $d$:
$q_{0\ldots00}, q_{0\ldots01}$ and so on.
The starting state is the state $q_{0\ldots 00}$
corresponding to the all-0 string.
The transition function is defined by
\[ \delta(q_{x_1\ldots x_d}, a, q_{y_1\ldots y_d})=
\prod_{j=1}^d \delta(q_{x_1}, a, q_{y_1}).\]
It is easy to see that this is just the tensor product of $d$ copies of $U_k$.
The automaton also has one accepting state and $2^d-1$ rejecting states.
After reading the right endmarker, the automaton passes to the accepting state
from $q_{0\ldots 00}$ and to a rejecting state from any other 
state $q_{x_1\ldots x_d}$. (To ensure unitarity, one-to-one correspondence
between $q_{x_1\ldots x_d}$ and rejecting states is established.)
A counterpart of Lemma \ref{L3} is

\begin{Lemma}
The state of $U'_k$ after reading $a^j$ is
\[ \underbrace{
(\cos\left(\frac{2\pi j k}{p}\right)|q_0\rangle
+i\sin\left(\frac{2\pi j k}{p}\right)|q_1\rangle)
\otimes\ldots\otimes (\cos\left(\frac{2\pi j k}{p}\right)|q_0\rangle
+i\sin\left(\frac{2\pi j k}{p}\right)|q_1\rangle)}_{
\mbox{$d$ times}}.\]
\end{Lemma}

The amplitude of $|q_0\rangle\otimes\ldots\otimes|q_0\rangle=q_{0\ldots 0}$
in this superposition is $\cos^d(\frac{2\pi j k}{p})$.
If $j$ is a multiple of $p$, then this is 1, meaning that 
words in $L_p$ are always accepted.
For $a^j\notin L_p$, we call $U'_k$ $\delta$-good if
it rejects $a^j$ with probability at least $1-\delta$.
We formulate a counterpart of Lemma \ref{L4}.

\begin{Lemma}
\label{L6}
For a suitable constant $d$, at least $1-\delta$ of all $U'_k$
are $\delta$-good.
\end{Lemma}

Then, we define a $\delta$-good sequence of automata as a sequence
such that, for any $a^j\notin L$, at least $1-2\delta$ of all automata 
in the sequence are $\delta$-good. 
Similarly to Lemma \ref{L5}, we show that there is a $\delta$-good sequence
$U'_{k_1}$, $U'_{k_2}$, $\ldots$ of length $O(\log n)$.
Then, we consider an automaton consisting of $U'_{k_1}$, $U'_{k_2}$, $\ldots$
and a distinguished starting state.
Upon reading the left endmarker $\cents$, it passes from the starting state
to a superposition where $|q_0\rangle$ states of all $U'_{k_l}$ have 
equal amplitudes. 
Again, it accepts $a^j\in L_p$ with probability 1 because all $U'_k$ accept
$a^j\in L_p$. Words $a^j\notin L_p$ are rejected by at least $1-2\delta$
of $U'_{k_l}$ with probability $1-\delta$. 
Therefore, the probability of rejecting any $a^j\notin L_p$ is
at least $(1-2\delta)(1-\delta) > 1-3\delta$.
Taking $\delta=\epsilon/3$ and choosing $d$ so that it
satisfies Lemma \ref{L6} completes the proof.
$\Box$

We have shown an exponential gap between deterministic 
and quantum 1-way finite automata. 
Next, we compare quantum and probabilistic finite automata.
Generally, probabilistic finite automata can recognize
some languages with the number of states being close to the logarithm
of the number of states needed by a deterministic automaton\cite{Ambainis,Freivalds}.
However, this is not the case with $L_p$.
Here, adding probabilism does not help to decrease the number of states at all.

\begin{Theorem}
\label{T8}
Any 1-way probabilistic finite automaton recognizing $L_p$ 
with probability $1/2+\epsilon$, for a fixed $\epsilon>0$, has
at least $p$ states.
\end{Theorem}

\noindent
{\bf Proof.}
Assume that there is a 1-way probabilistic finite automaton
with less than $p$ states recognizing $L_p$ with probability
$\frac{1}{2}+\epsilon $, for a fixed
$\epsilon > 0$. Since the language $L_p$ is in a single-letter alphabet,
the automaton can be described as a Markov chain. We use the
classification of Markov chains described in Section 2 of
\cite{KemenySnell60} . According to this classification, the states of the
Markov chain (the automaton) are divided into ergodic and transient
states. An ergodic set of states is a set which cannot be left once it is
entered. A transient set of states is a set in which every state can be
reached from every other state, and which can be left. An ergodic state is
an element of an ergodic set. A transient state is an element of a
transient set.

If a Markov chain has more than one ergodic set, then there is absolutely
no interaction between these sets. Hence we have two or more unrelated
Markov chains lumped together. These chains may be studied seperately. If
a Markov chain consists of a single ergodic set, the chain is called an
ergodic chain. According to results in Section 2 of \cite{KemenySnell60} ,
every ergodic chain is either regular or cyclic.

If a Markov chain is regular, then sufficiently high powers of the state
transition matrix $P$ of the Markov chain are with all positive elements.
Thus no matter where the process starts, after sufficient lapse of time it
can be in any state. Moreover, by Theorem 4.2.1 of \cite{KemenySnell60} 
there is limiting vector of probabilities of being in the states of the
chain, not dependent of the initial state.

If a Markov chain is cyclic, then the chain has a period $d$, and its
states are subdivided into $d$ cyclic sets $(d > 1)$. For a given starting
position, it moves through the cyclic sets in a definite order, returning
to the set of the starting state after $d$ steps. Hence the $d$-th power
of the state transition matrix $P$ describes a regular Markov chain.

We have assumed that $p$ is prime, and the automaton has less than $p$
states. Hence for every cyclic state of the automaton the value of $d$ is
strictly less than $p$, and because of primality of $p$, $d$ is relatively
prime to $p$. By $D$ we denote the least common multiple of all such
values $d$. Hence $D$ is relatively prime to $p$, and so is any positive
degree $D^n$ of $D$. Since $1^{D^n} \notin M_p$ but $1^{D^n \, p} \in
M_p$, the total of the probabilities to be in an accepting state exceeds
$\frac{1}{2}+\epsilon $ for $1^{D^n}$ and is less than
$\frac{1}{2}- \epsilon $ for $1^{D^n \, p}$. Contradiction with
Theorem 4.2.1 of \cite{KemenySnell60}.
$\Box$

\begin{Corollary}
For the language $L_p$, the number of states needed by a
classical (deterministic or probabilistic) 1-way
automaton is exponential in the number of states of a 1-way QFA.
\end{Corollary}

\noindent
{\bf Proof:}
Follows from Theorems \ref{T6} and \ref{T8}.
$\Box$

\subsection{Equality}
\label{S42}

Divisibility by a prime is quite natural problem and we expect that
our algorithm can be used as a subroutine, making other quantum 
algorithms more space-efficient.
Here, we show how to use our quantum automaton for another problem 
as well. This problem is checking whether the length of the input
word is equal to some constant $n$.

%Now, we proceed more formally.
%Let $L'_n$ be a language consisting of one word $a^n$ 
%in a single-letter alphabet. Then, the following is true.

\begin{Theorem}
\label{T10}
\cite{Freivalds}
Let $L'_n$ be a language consisting of one word $a^n$ 
in a single-letter alphabet.
\begin{enumerate}
\item
Any deterministic automaton that recognizes $L'_n$ has at least $n$ states.
\item
For any $\epsilon>0$, there is a probabilistic automaton 
with $O(\log^2 n)$ states recognizing $L'_n$ with probability $1-\epsilon$.
\end{enumerate}
\end{Theorem}

The first part is evident. To prove the second part, Freivalds\cite{Freivalds} 
used the following construction.
$O (\frac{\log n}{\log \log n})$
different primes are employed and $O (\log n)$ states are used for every
employed prime. At first,
the automaton randomly chooses a prime $p$, and then the remainder modulo
$p$ of the length of the input word is found and compared with the
standard. %However the count is not completely deterministic. 
Additionally, once in every $p$ steps
a transition to a rejecting state is made with a "small" probability
$\frac{const \, p}{n}$. The number of used primes suffices to assert
that, for every input of length less than $n$,  most of primes $p$ give
remainders different from the remainder of $n$ modulo $p$. The "small"
probability is chosen to have the rejection probability high enough for
every input length $N$ such that both $N \ne n$ and an $\epsilon
$-fraction of all the primes used have the same remainders $mod \, p$ as $n$.

This 1-way probabilistic automaton is reversible according to the definition
of section \ref{S2}. We can use Theorem \ref{T9} to transform it
into quantum automaton with the number of states increasing at most twice.
Then, we obtain a counterpart of Theorem \ref{T10} for quantum case.

%The construction of the probabilistic automaton in Theorem \ref{T10}
%is based on choosing a prime randomly from some set and counting
%modulo this prime. We can plug in the space-efficient 1-way QFA 
%from Theorem \ref{T6} and decrease the number of states 
%(see \cite{AF} for details). This gives us

However, we can do better by counting modulo prime as in Theorem \ref{T6}.
For that, we need $O(\log p)$ states for each prime $p$
(instead of $p$ states in the probabilistic case).
Each prime $p$ is $O(\log n)$ and there are $O(\log n/\log\log n)$ of
them. Therefore, the number of states in quantum case will be 
\[ O(\frac{\log n}{\log\log n}\log p) = O(\frac{\log n}{\log\log n}\log \log n)=
O(\log n) .\]
We have shown

\begin{Theorem}
$L'_n$ can be recognized by a 1-way QFA with $O(\log n)$ states.
\end{Theorem}

Again, the QFA is exponentially smaller than the corresponding
deterministic automaton.

\subsection{Are QFAs always space-efficient?}
\label{S43}

Subsections \ref{S41} and \ref{S42} showed cases when 1-way QFAs are more
space-efficient than their classical counterparts.
There can be examples of different kind where 
deterministic finite automata are exponentially smaller than 1-way
QFAs. The construction of theorem \ref{T2} which transforms the 
minimal automaton into a 1-way RFA can increase
the size of the automaton exponentially.
The next theorem shows that this is inevitable.

\begin{Theorem}
\label{T11}
Let $L_m=(xy|zy)^m\cup\{(xy|zy)^i xx|0\leq i\leq m-1\}$. 
Then,
\begin{enumerate}
\item
$L_m$ can be recognized by a 1-way deterministic 
finite automaton with $3m+2$ states;
\item
$L_m$ can be recognized by a 1-way reversible automaton but it
requires at least $3(2^m-1)$ states.
\end{enumerate}
\end{Theorem}

%\noindent
%{\bf Proof.}
%We use $L_m=(xy|zy)^m\cup\{(xy|zy)^i xx|0\leq i\leq m-1\}$.
%Details are omitted in this version.
%$\Box$ 

After first version of this paper appeared, Ambainis, Nayak
and Vazirani\cite{ANV} showed that, for a different
language, the number of states needed by a 1-way QFA is
almost exponentially bigger than the number of states
of a 1-way deterministic finite automaton. 

%It remains open whether the same is true for 1-way quantum automata (instead
%of reversible automata) or quantum effects can be used to make automata
%more space-efficient in this case as well.

\section{Modifications of 2-way QFAs}

%1-way and 2-way QFAs have different advantages and disadvantages. 
%the 2-way model is unrealistic
%and 1-way model is much more reasonable. 
The advantage of 1-way quantum automata is the simplicity of this model.
%are much simpler than 2-way automata.
%In the sequel
However, we saw that 1-way automata are quite limited in several
situations (despite being good in others) while \cite{KW} shows
that 2-way QFAs are strictly more powerful than classical finite automata.
It would be interesting to come up with a model having both advantages, i.e.
being both powerful and simple.
In the remainder, we propose several modifications of quantum
automata which are intermediate between 1-way QFAs and 2-way QFAs.
Quantum part is kept finite in all of these models. %and representing 
%input in a quantum form is not necessary in any of them.
Questions about exact power of these models %, relations between them and
%their classical counterparts 
are mostly open but we have shown that,
in most of these models, all regular languages can be recognized and,
in at least one of them, non-regular languages can be recognized as well.

\subsection{Scanning the tape multiple times}
\label{S51}

The simplest modification is to allow a 1-way QFA 
to scan its input tape several times
(after the right endmarker it goes to the left endmarker and so on).
This is enough to make the proof from \cite{KW} that 1-way QFAs 
recognize only regular languages fail. 
If we allow the automaton to reject words by non-halting,
a nonregular language can be recognized.

\begin{Theorem}
\label{T11a}
Let 
$L=\{ a^nb^n |n\in\bbbn\}$.
There is a 1-way QFA $M$ scanning tape several times such that
\begin{enumerate}
\item
If $x\notin L$, $M$ stops with probability 1 after $O(|x|)$ scans of the tape.
\item
If $x\in L$, $M$ never stops.
\end{enumerate}
\end{Theorem}
 
If we require $M$ to stop in a rejecting state for rejection, a similar
question is still open. It is also open whether multiple scans of the tape 
can be used by a 1-way QFA to recognize an arbitrary regular language.
(However, known proofs that 1-way QFAs do not recognize some regular languages
also fail in this case.)

\subsection{Passing information back to environment}
\label{S52}

%A QFA says to the environment after each step whether it has halted
%(from its observation). Why it cannot say us more?
Another possibility is introducing more complicated observables.
We can partition all non-halting states into 2 or 3 classes:
moving-left states, moving-right states and (may be) non-moving states.
Then, after each step we observe whether the automaton is in accepting,
rejecting, moving-right or moving-left state.
If it is in a halting state, we terminate the computation.
If it is in a moving-right state, we feed it the next letter
(do the transformation on the quantum system corresponding to
the next letter). If it is in a moving-left state, we feed it the 
previous letter. 

The model of section \ref{S51} is a special case of this model 
where all non-halting states are classified as moving-right states.

\subsection{Preprocessing the input word}
\label{S53}

In this model, we have two automata $M_1$ and $M_2$ instead of one.
$M_1$ is a 2-way deterministic (or probabilistic) finite automaton
with output and $M_2$ is a 1-way QFA. 
The input word is given to $M_1$ and $M_2$ is run on the output of
$M_1$. (This can be viewed as $M_1$ preprocessing the input word.)
Again, the model of section \ref{S51} can be viewed as a special
case of this model where $M_1$ moves from left to right all the time
and outputs all letters that it reads.

Any regular language can be recognized in a trivial way because 
we can recognize it by $M_1$ and give the result as an input to $M_2$.
If the preprocessing is done by a probabilistic automaton,
we can do more.

\begin{Theorem}
\label{T12}
For any $\epsilon>0$, there is a 2-way probabilistic finite automaton $M_1$
and a 1-way QFA $M_2$ such that, with probability at least $1-\epsilon$,
\begin{enumerate}
\item
$M_1$ stops in time quadratic in the length of the input and,
\item
$M_2$ accepts the output of $M_1$ if and only if $x\in\{ a^nb^n |n\in\bbbn\}$.
\end{enumerate}
\end{Theorem}

Any 2-way probabilistic automaton that recognizes a non-regular language
has an exponential expected running time\cite{DS89,Fr81,GW86,KF91}.
So, neither polynomial time 2-way probabilistic finite automata nor 1-way QFAs
can recognize non-regular languages. 
However, their combination can do that! 

\medskip
\noindent
{\bf Acknowledgments.}
We thank Isaac Chuang, Ashwin Nayak, Alistair Sinclair,
Amnon Ta-Shma and Umesh Vazirani for
useful discussions.

\end{document}